\documentstyle[aps,prb,epsfig]{revtex}
\begin{document}
\baselineskip=0.5cm
\renewcommand{\thefigure}{\arabic{figure}}
\title{Self-consistent scattering theory of the pair distribution function
in charged Bose fluids}
\author{B. Davoudi$^{1,2}$, R. Asgari$^{1,2}$, M. Polini$^1$ and M. P. Tosi$^1$}
\address{$^1$NEST-INFM and Classe di Scienze, Scuola Normale Superiore, I-56126 Pisa, Italy\\ 
$^2$Institute for Studies in Theoretical Physics and Mathematics, Tehran, P.O.Box 19395-5531,Iran\\
}
\maketitle
\begin{abstract}
We use a density functional theoretical approach to calculate the pair distribution function 
and the effective interactions in homogeneous fluids of spinless charged bosons. The scheme 
involves the self-consistent solution of a two-particle scattering problem with an effective 
scattering potential which embodies many-body effects and is adjusted to the compressibility 
sum rule. Numerical results are presented over an extensive range of density in both three and 
two dimensions.
\end{abstract}
\vspace{0.5 cm}

Exchange and correlations in fluids of charged particles have been a focus of interest in 
many-body physics for many decades. An important manifestation of these effects is the 
equilibrium distribution $g(r)$ of pairs of particles in the ground state of the homogeneous fluid, 
which describes its state of short-range order. Knowledge of $g(r)$ as a function of the fluid 
density determines its energy and is essential in the construction of energy functionals for 
applications of density functional theory (DFT) beyond the local density approximation~\cite{1}. 
Main attention has been given to electron fluids because of their relevance in the physics of 
metals and semiconductors, but there also is an interest for plasmas of charged bosons in 
quantum statistical mechanics, for instance in regard to condensates of point-like Cooper pairs 
as models for superfluid states~\cite{4} or to the equation of state and nuclear reactions in 
astrophysical plasmas~\cite{5}.

	There has recently been a renewed interest in the study of $g(r)$ for electron gas models 
within a two-body scattering approach stemming from work by Overhauser~\cite{6}. In brief, $g(r)$ is 
obtained from the solution of a Schršdinger equation for particle-pair wave functions with 
effective scattering potentials which, starting from a simple electrostatic model~\cite{6,7}, have been 
developed into a self-consistent Hartree model~\cite{8} and into spin-dependent effective pair 
interactions~\cite{9}. In the present work we derive a DFT basis for such an approach and, using an 
earlier modelling of the effective interactions to incorporate the thermodynamic sum rules~\cite{10}, 
we develop a fully self-contained and self-consistent determination of $g(r)$ and of the effective 
scattering potential. Our focus here is on plasmas of spinless Bose particles, leaving aside the 
further complications that arise in the presence of the spin degree of freedom as already briefly 
discussed elsewhere~\cite{10,11}.

	We consider, therefore, a quantum fluid of point-like bosons having charge $e$ and mass $m$, 
which are free to move in dimensionality $D = 2$ or $3$ at zero temperature and are neutralized by a 
uniformly charged background of density $n$. This model will be referred to as a charged Bose 
fluid (CBF). Our aim is to use DFT for building a self-consistent theory that gives the pair 
distribution function $g(r)$ as output. From the Hohenberg-Kohn theorem~\cite{12} the ground state 
energy of the fluid in the presence of an external potential $V_{\rm ext}({\bf r})$ can be written as a functional of the density profile $n({\bf r})$ in the form
\begin{equation}\label{e1}
E_{\rm gs}[n({\bf r})]=T_s[n({\bf r})]+E_{\scriptscriptstyle \rm H}[n({\bf r})]+\int\,d^D{\bf r}\,\,V_{\rm ext}({\bf r})\Delta n({\bf r})+E_{\rm c}[n({\bf r})]
\end{equation}
where $\Delta n({\bf r})=n({\bf r})-n$ and $T_s$ is the ideal kinetic energy functional. The Hartree term  $E_{\scriptscriptstyle \rm H}$ is given by
\begin{equation}\label{e2}
E_{\rm H}[n({\bf r})]=\frac{1}{2}\int d^D{\bf r}\int d^D{\bf r}'\,v(|{\bf r}-{\bf r}'|)\,\Delta n({\bf r})\Delta n({\bf r}')
\end{equation}
where $v(|{\bf r}-{\bf r}'|)=e^2/|{\bf r}-{\bf r}'|$. 
The last term in Eq.~(\ref{e1}) is the correlation energy functional, which 
contains all the quantum many-body (QMB) effects. 
In Eqs.~(\ref{e1}) and~(\ref{e2}) the presence of a 
neutralizing background has been taken into account.

The quantity $n[g(r)-1]$ in the homogeneous fluid can be viewed as the distortion in the 
density profile that instantaneously surrounds a particle of the fluid located at position ${\bf r}=0$~\cite{13}. As usual $g(r)$ is defined by setting $ng(r)\Omega_D r^{D-1} dr$, with  $\Omega_2 = 2\pi$ 
and $\Omega_3= 4\pi$, equal to the average number of particles inside a shell of radius $r$ and thickness $dr$ centered on the particle at the origin. 
The appropriate ground-state energy functional for the surrounding fluid is obtained 
from Eq.~(\ref{e1}) with the formal replacements $V_{\rm ext}({\bf r}) \rightarrow v(r)$ and
\begin{equation}\label{e3}
\Delta n({\bf r})=n[g(r)-1]~.
\end{equation}  
Following the treatment given for inhomogeneous fluids in the book of Dreizler and Gross~\cite{12}, 
a formal expression for the QMB energy functional can be obtained via an adiabatic connection 
formula~\cite{12},
\begin{equation}\label{e4}
E_{\rm c}[n({\bf r})]=\frac{1}{2}\frac{1}{e^2}\int_{0}^{e^2} d \lambda
\int d^D{\bf r}\int d^D{\bf r}'\,v(|{\bf r}-{\bf r}'|)~n({\bf r})n({\bf r}')~ \left[g_{\lambda}^{(3)}({\bf r}, {\bf r}')-1\right]
\end{equation}                                       
where the integration over the coupling strength $\lambda$ accounts for the shift in kinetic energy that 
accompanies the switching on of the interactions. In Eq.~(\ref{e4}) we have $n({\bf r})=ng(r)$ and $g_{\lambda}^{(3)}({\bf r}, {\bf r}')$ measures the probability of finding two particles at ${\bf r}$ and ${\bf r}'$ when a third particle is at the origin, the interaction potential being $v_{\lambda}(|{\bf r}-{\bf r}'|)=\lambda\,e^2/|{\bf r}-{\bf r}'|$. Of course, $g_{\lambda}^{(3)}({\bf r}, {\bf r}')$  depends 
functionally on  $n({\bf r})$.

	An Euler-Lagrange equation for $g(r)$ in the boson fluid can now be obtained from the 
variational principle, using the ideal kinetic energy functional which for bosons is given by the von Weizs\"{a}cker expression~\cite{14}. This is
\begin{equation}\label{e5}
T_s[g(r)]=\frac{\hbar^2 n}{8 m_r}\,\int d^D{\bf r}\,\frac{|\nabla g(r)|^2}{g(r)}
\end{equation} 
where $m_r=m/2$ is the reduced mass of a particle pair. Taking the zero of energy at the 
chemical potential, the equation for $g(r)$ reads
\begin{equation}\label{e6}
\left[-\frac{\hbar^2}{2 m_r}\nabla^2_{\bf r}+V_{\scriptscriptstyle \rm KS}(r)\right]\sqrt{g(r)}=0
\end{equation}
where $V_{\scriptscriptstyle \rm KS}(r)$ is the Kohn-Sham scattering potential,
\begin{equation}\label{e7}
V_{\scriptscriptstyle \rm KS}(r)=v(r)+\int d^D{\bf r}' v(|{\bf r}-{\bf r}'|)\Delta n({\bf r}')+
\frac{\delta E_{\rm c}[n({\bf r})]}{\delta n({\bf r})}~.
\end{equation}
In fact, Eq.~(\ref{e6}) can also be obtained from the Kohn-Sham mapping in DFT~\cite{12} by building 
$\Delta n({\bf r})$ from Kohn-Sham scattering orbitals $\Phi_{\bf k}({\bf r})$. 
Since at zero temperature all bosons in the reference Kohn-Sham ideal gas are in the zero momentum state, we have $g(r)\propto |\Phi_{{\bf k}=0}({\bf r})|^2$ and Eq.~(\ref{e6}) is just the Kohn-Sham Schr\"odinger equation for the pair wave function $\Phi_{{\bf k}=0}({\bf r})$ at zero relative momentum. We conclude, therefore, that for a Bose fluid the scattering-theory approach 
to the pair distribution function admits a rigorous DFT derivation, which yields Eqs.~(\ref{e6}) and~(\ref{e7}).

	Of course, the functional dependence of the QMB energy on density is not known and at 
this point we have to resort to some approximations. Their goodness can only be gauged {\it a 
posteriori}. Firstly, the three-body correlation function  $g_{\lambda}^{(3)}({\bf r}, {\bf r}')$ in Eq.~(\ref{e4}) would lead us into a 
hierarchy of higher-order correlation functions. The simplest way of truncating this hierarchy is 
to replace $g_{\lambda}^{(3)}({\bf r}, {\bf r}')$ by  $g_{\lambda}(|{\bf r}- {\bf r}'|)$, in analogy with what has been done in treating the equation of motion for the Wigner distribution function in the electron gas subject to external potentials~\cite{16}. 
This approximation can be expected to work well at strong coupling where the probability 
of simultaneously finding three particles inside a radius $r_s a_B$ is small, such a "strong coupling" 
situation being reached at lower values of $r_s$ in 2D than in 3D. 
As a second approximation we expand the QMB energy in a functional Taylor series in powers of $\Delta n({\bf r})$ up to second order 
terms. With the definition
\begin{equation}\label{e8}
{\rm f}(|{\bf r}-{\bf r}'|)\equiv \left.\frac{\delta^{2} E_{\scriptscriptstyle \rm c}[n(r)]}{\delta n(r)\delta n(r')}\right|_{\Delta n(r)=0}
\end{equation}
we find in Fourier transform
\begin{equation}\label{e9}
V_{\scriptscriptstyle \rm KS}(q)=v(q)+v(q)[1-G(q)]~\Delta n(q)~.
\end{equation}
Here, $G(q)\equiv -f(q)/v(q)$ is the so-called local field factor, defined in terms of the Fourier transform $f(q)$ of the QMB kernel in Eq~(\ref{e8}), and $v(q)$ is the Fourier transform of the Coulomb potential ({\it i.e.} $v(q)= 4 \pi e^2/q^2$ in $D=3$, $v(q)= 2 \pi e^2/q$ in $D=2$). Finally, $\Delta n(q)$ in Eq.~(\ref{e9}), is given by
\begin{equation}\label{e10}
\Delta n(q)=S(q)-1
\end{equation}
where $S(q)$ is the structure factor, related to the pair function by
\begin{equation}\label{e11}
S(q)=1+n\,\int d^{D} {\bf r}\,[g(r)-1]\exp{(-i {\bf q} \cdot {\bf r})}~.
\end{equation}
The solution of Eqs.~(\ref{e6}) and (\ref{e9}) can therefore be carried out self-consistently, given knowledge of the local-field factor $G(q)$. The results of such a self-consistent scheme (SCS) for the 3D-
CBF, using the data on $G(q)$ from the Quantum Monte Carlo (QMC) study of Moroni {\it et al.}~\cite{17}, will be reported in Figures~\ref{f1} and \ref{f2} below.

	In fact, the present approach can be extended into a fully self-contained theory in which the 
local field factor is self-consistently determined over the relevant $q$-range during the calculation 
rather than taken as input from QMC data. In such a fully self-consistent scheme (FSCS) we 
adopt a closure relation between  $G(q)$ and $S(q)$, 
which satisfies the compressibility sum rule~\cite{10}. 
That is, we set
\begin{equation}\label{e12}
G(q)=D_n\,{\mathcal G}(q)
\end{equation}
where the differential operator $D_n$ is defined by
\begin{equation}\label{e13}
D_n\equiv 1+2n\frac{\partial}{\partial n}+\frac{1}{2}n^2\,
\frac{\partial^2}{\partial n^2}~
\end{equation}
while ${\mathcal G}(q)$ is given by 
\begin{equation}\label{e14}
{\mathcal G}(q)\equiv -\frac{1}{n v(q)}\,\frac{1}{e^2}\,\int_{0}^{e^2} d\lambda\, \int \frac{d^{D} {\bf q}'}{(2 \pi)^D}\,v(q')~\left[S_{\lambda}(|{\bf q}+{\bf q}'|)-1\right]~,
\end{equation}
with $S_{\lambda}(q)$ being the partial structure factor at coupling constant $\lambda$. 
Although these expressions 
are strictly correct only in the long-wavelength limit, they yield a good account of QMC data on 
local field factors over the relevant range of $q$ (see Ref.~8 and Figure~\ref{f4} below). 
The explicit proof that Eqs.~(\ref{e12})-(\ref{e14}) 
satisfy the compressibility sum rule at long wavelengths requires two 
simple steps: (i) the dependence of the integral in Eq.~(\ref{e14}) on wave number $q$ can be neglected in 
the limit $q \rightarrow 0$, yielding ${\mathcal G}(q) \rightarrow -2 \varepsilon_{\rm c}(n)/[n v(q)]$  in this limit where $\varepsilon_{\rm c}(n)$ is the correlation 
energy of the CBF; and hence (ii) $D_n {\mathcal G}(q)\rightarrow -1/[n^2 \kappa\, v(q)]$ with $\kappa$ being the compressibility of the CBF.

\begin{figure*}
\begin{center}
\includegraphics[scale=0.6]{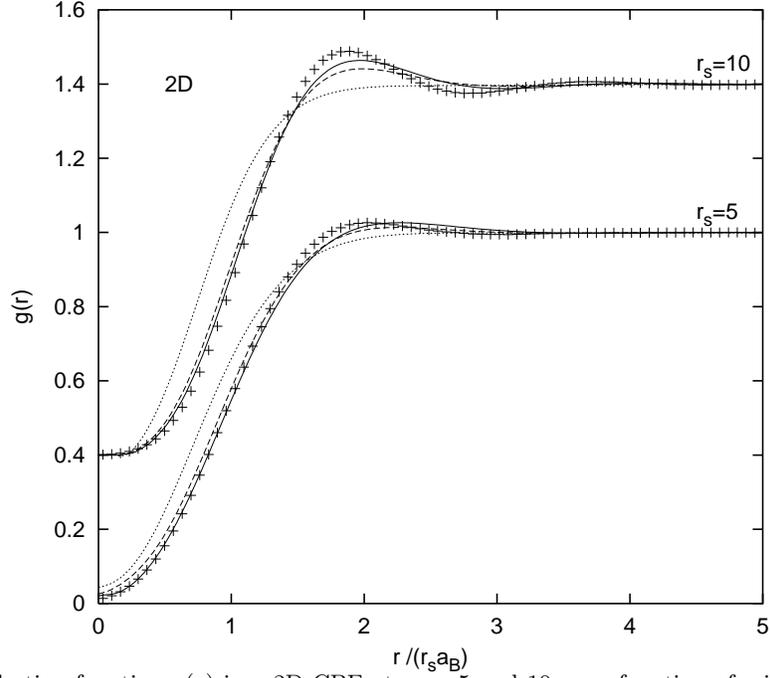}
\caption{The pair distribution function $g(r)$ in a 2D-CBF at  $r_s=5$ and $10$, as a function of $r$ in 
units of $r_s a_B$. The results of the FSCS (full lines), of the HA (dotted lines) and of the HNC/EL 
theory (dashed lines) are compared with QMC data of Dr. S. Moroni (crosses, unpublished). 
The curves at $r_s=10$ have been shifted upwards by $0.4$.}
\label{f3}
\end{center}
\end{figure*}

We turn at this point to a presentation of our numerical result. We have solved the FSCS 
based on Eqs.~(\ref{e6}), (\ref{e9})-(\ref{e11}) and (\ref{e12})-(\ref{e14}) 
for a 3D-CBF with coupling strength up to $r_s=20$ (with $r_s a_B= (4 \pi n/3)^{-1/3}$) and for a 2D-CBF with coupling strength up to $r_s=10$ (with $r_s a_B=(\pi n)^{-1/2}$ ). The main results of our work are shown in Figures~\ref{f1}-\ref{f4}.

	In Figure~\ref{f1} we show that our FSCS results for $g(r)$ in the 3D-CBF at $r_s=10$ and $r_s=20$ 
are in excellent agreement with the QMC data of Moroni {\it et al.}~\cite{17}. 
We also compare our results with those that we have obtained in the Hartree approximation (HA, setting $G(q)=0$) and with those that we have recalculated by the hypernetted-chain Euler-Lagrange approach 
(HNC/EL) of Apaja {\it et al.}~\cite{18}, the latter being still based on Eq.~(\ref{e6}) 
but with a different choice for the dependence of the scattering potential on $S(q)$. 
The HA is unable to reproduce the 
emergence of a first-neighbor shell with increasing coupling strength, as already noted in its use 
for the study of pair correlations in the electron gas~\cite{8}. The two self-consistent approaches to 
the dependence of the scattering potential on the structure factor, on the other hand, are in very 
good agreement with each other.

	The inset in Figure~\ref{f1} shows that our SCS results at $r_s=20$ are in very good agreement 
with those obtained in the FSCS, stressing that our self-consistent determination of $G(q)$ from 
the compressibility sum rule also accounts with sufficient accuracy for this function over the 
relevant range of wave number. We have made use of this observation to test the accuracy of the 
present theory for $g(r)$ at very strong coupling, as is shown in Figure~\ref{f2} by reporting SCS results for the 
3D-CBF at $r_s=100$ in comparison again with the QMC data of Moroni {\it et al.}~\cite{17} on $g(r)$ and 
with the results of the HA and of the HNC/EL~\cite{18}.

In Figure~\ref{f3} we illustrate the quality of our results for the 2D-CBF at $r_s=5$ and $r_s=10$. 
No data are as yet available for the local-field factor in this model system, and we show in this 
Figure our FSCS results in comparison with QMC data on $g(r)$ kindly sent to us by Dr. Moroni 
prior to publication. Again we obtain excellent agreement with the QMC data and with the 
HNC/EL theory of Apaja {\it et al.}~\cite{18}.

Finally, in Figure~\ref{f4} we show our FSCS results for the local-field factor $G(q)$ in the 3D-CBF at  $r_s=10$, in comparison with the QMC data of Moroni {\it et al.}~\cite{17}, and for the 2D-CBF at $r_s=5$. Deviations from the QMC data are seen to emerge in $3D$ with increasing wave number, starting 
at $q r_s a_B>1.5$ and becoming very large at $q r_s a_B>4$. However, as already stressed in the 
discussion of the results for $g(r)$ in Figure~\ref{f1}, 
these discrepancies are scarcely of any relevance in the calculation of pair correlations once the compressibility sum rule is embodied into the theory.

	In summary, we have shown that a scattering-theory approach to pair correlations in boson 
plasmas has a sound theoretical justification within the framework of DFT and leads to fully 
quantitative results when self-consistence between pair correlations and effective particle-particle 
interactions is incorporated into the theory. It may be worth exploring in the future alternative 
approximations to the truncated expansion of the QMB energy functional.

	As a final remark we wish to point out that, having taken the Kohn-Sham viewpoint in our 
choice of an ideal Bose gas as the reference DFT fluid, our approach carries no information on 
momentum distributions and in particular on the depletion of the condensate which occurs with 
increasing coupling strength. An extended version of the DFT scheme, which adopts both the 
particle density and the order parameter of the condensate as basic variables of the 
inhomogeneous fluid and uses the Bogoliubov - de Gennes equations as reference, has been 
developed for such wider purposes~\cite{19}.

\acknowledgements
This work was partially funded by MIUR under the PRIN2001 Initiative and by INFM 
under the PRA2001 Program. We are indebted to Dr. S. Moroni for providing us with 
his unpublished QMC data reported in Figure~\ref{f3}.
\newpage

\begin{figure*}
\begin{center}
\includegraphics[scale=0.6]{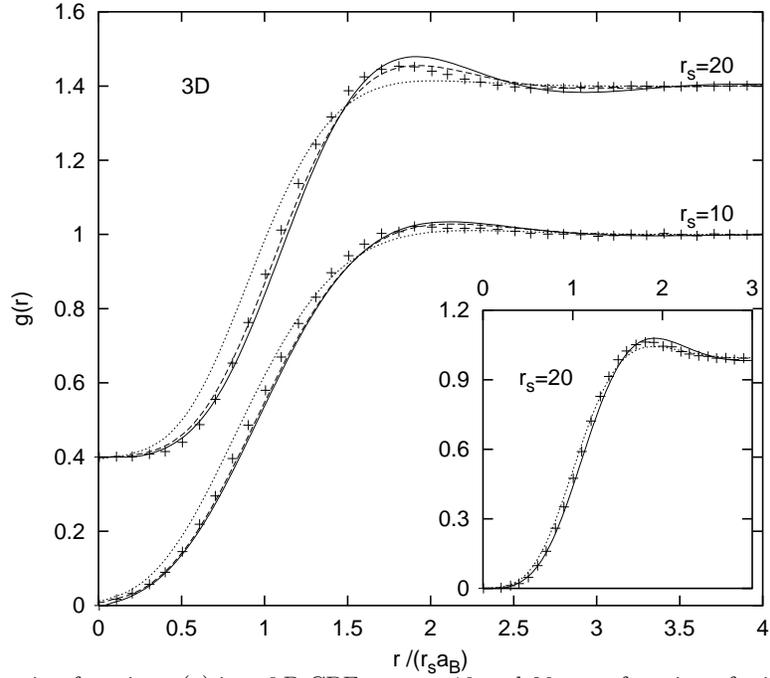}
\caption{The pair distribution function $g(r)$ in a $3D$-CBF at $r_s=10$ and $20$, as a function of $r$ in units of $r_s a_B$. In the main body of the Figure the results of the FSCS (full lines), of the HA (dotted lines) and of the HNC/EL theory (dashed lines) are compared with QMC data from Ref.~14 (crosses). The curves at $r_s=20$ have been shifted upwards by 0.4. In the inset the SCS results at $r_s=20$ (dotted line) are compared with the FSCS results (solid line) and with the QMC data (crosses). 
}
\label{f1}
\end{center}
\end{figure*}
\begin{figure*}
\begin{center}
\includegraphics[scale=0.6]{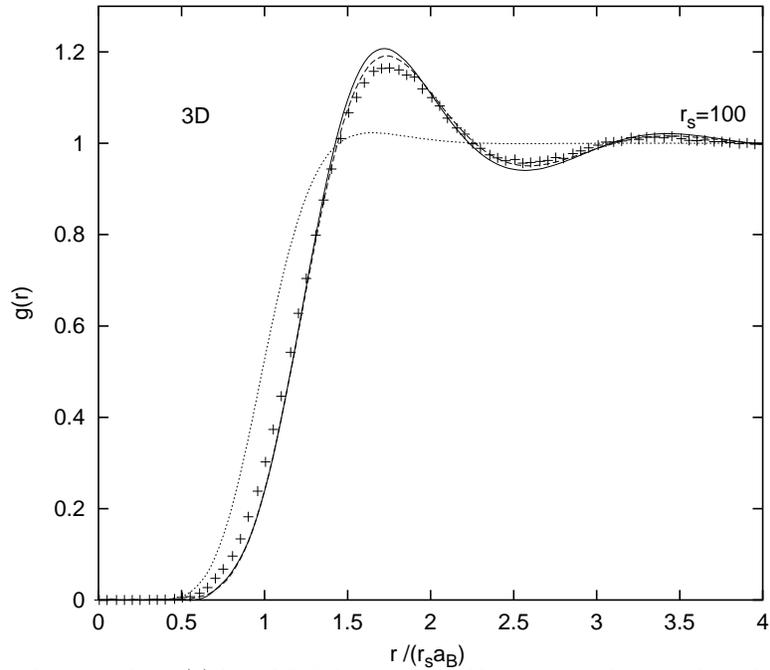}
\caption{The pair distribution function $g(r)$ in a 3D-CBF at $r_s=100$, as a function of $r$ in units 
of 	$r_s a_B$. The results of the SCS (full line), of the HA (dotted line) and of the HNC/EL theory 
(dashed line) are compared with QMC data from Ref.~14 (crosses).}
\label{f2}
\end{center}
\end{figure*}
\begin{figure*}
\begin{center}
\includegraphics[scale=0.6]{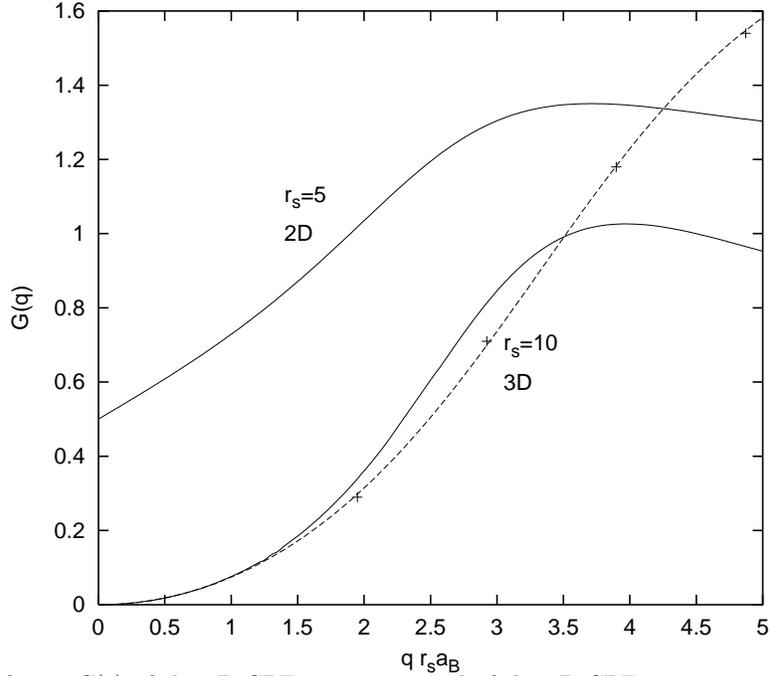}
\caption{The local-field factor $G(q)$ of the 3D-CBF at $r_s=10$ and of the 2D-CBF at $r_s=5$, as 
a function of $q$ in units of $(r_s a_B)^{-1}$. The FSCS results are shown as full lines and are compared 
for the 3D-CBF with QMC data from Ref.~14 (crosses) and with an interpolation formula to the QMC results reported in Ref.~14 (dashed line). The curve giving the FSCS results 
for the 2D-CBF has been shifted upwards by $0.5$.}
\label{f4}
\end{center}
\end{figure*}

\end{document}